\documentclass[a4paper,12pt]{article}
\usepackage{blindtext}
\usepackage[T1]{fontenc}
\usepackage{lmodern}
\usepackage[utf8]{inputenc}
\usepackage{mathrsfs}
\usepackage{epsfig}
\usepackage{epstopdf}
\usepackage{bigints}
\usepackage{lastpage}
\usepackage{comment}
\usepackage{setspace}
\usepackage{enumerate}
\usepackage{amsmath}
\usepackage{float}
\usepackage{array}
\usepackage{dcolumn}
\usepackage[top=1in, bottom=1in, left=0.7in, right=0.7in]{geometry}
\usepackage{multirow}
\usepackage[table,xcdraw]{xcolor}
\usepackage{lscape}
\usepackage{breqn}

\usepackage{lipsum}
\usepackage{color}
\usepackage{setspace}

\usepackage{framed}
\usepackage{amsthm}
\usepackage{bm}
\usepackage{mathrsfs}
\usepackage{amssymb}
\usepackage{relsize}
\usepackage{graphicx}
\usepackage{algorithmicx}
\usepackage{algorithm}
\usepackage{algpseudocode}
\usepackage{abstract}
\usepackage{natbib}
\usepackage{booktabs}
\usepackage{hyperref}

\usepackage{pdflscape}
\usepackage{siunitx}

\usepackage[flushleft]{threeparttable}

\newcommand{\E}{\mathop{\mbox{\sf E}}}

\def\E{\mathbb{E}}

\usepackage{natbib}
\bibpunct{(}{)}{;}{}{,}{,}
\usepackage{xcolor}
\hypersetup{
    colorlinks,
    linkcolor={red!50!black},
    citecolor={blue!50!black},
    urlcolor={blue!80!black}
}

\theoremstyle{definition}

\newtheorem{theorem}{Theorem}[section]
\newtheorem{lemma}[theorem]{Lemma}
\newtheorem{proposition}[theorem]{Proposition}

\newcommand{\Rplus}{\protect\hspace{-.1em}\protect\raisebox{.35ex}{\smaller{\smaller\textbf{+}}}}
\newcommand{\Cpp}{\mbox{C\Rplus\Rplus}\xspace}

\title{\textbf{Sentiment-Driven Stochastic Volatility Model: \\ A High-Frequency Textual Tool for Economists}\footnote{Financial support from the Deutsche Forschungsgemeinschaft via IRTG 1792 ``High Dimensional Non Stationary Time Series,''  Humboldt-Universit\"{a}t zu Berlin, and the Czech Science Foundation under the 19-28231X (EXPRO) project is gratefully acknowledged.}}

\author{Jozef Barunik \thanks{Institute of Economic Studies, Charles University, Opletalova 26, 110 00, Prague, CR and Institute of Information Theory and Automation, Academy of Sciences of the Czech Republic, Pod Vodarenskou Vezi 4, 182 08, Prague, Czech Republic (\textit{e-mail: \href{mailto:barunik@fsv.cuni.cz}{barunik@fsv.cuni.cz}})(corresponding author) 
 }
 \and
Cathy Yi-Hsuan Chen \thanks{Adam Smith Business School, University of Glasgow, UK (\textit{e-mail: \href{mailto:CathyYi-Hsuan.Chen@glasgow.ac.uk}{CathyYi-Hsuan.Chen@glasgow.ac.uk}}) }
\and 
Jan Vecer \thanks{Charles University in Prague - Faculty of Mathematics and Physics, Sokolovska 83, Prague, 186 75 Czech Republic (\textit{e-mail: \href{mailto:vecer@karlin.mff.cuni.cz}{vecer@karlin.mff.cuni.cz}})}}

\begin{document}

\maketitle 
 \thispagestyle{empty}

\newcolumntype{d}[1]{D{.}{.}{#1}}

\newcolumntype{.}{D{.}{.}{-1}}

\begin{abstract}

\noindent  

We propose how to quantify high-frequency market sentiment using high-frequency news from \href{http://www.NASDAQ.com/news}{NASDAQ news platform} and support vector machine classifiers. News arrive at markets randomly and the resulting news sentiment behaves like a stochastic process. To characterize the joint evolution of sentiment, price, and volatility, we introduce a unified continuous-time sentiment-driven stochastic volatility model. We provide closed-form formulas for moments of the volatility and news sentiment processes and study the news impact. Further, we implement a simulation-based method to calibrate the parameters. Empirically, we document that news sentiment raises the threshold of volatility reversion, sustaining high market volatility.  \\

\noindent


\end{abstract}

\setcounter{section}{0}

\section{Introduction}
\doublespacing

Sentiment measures based on textual analysis of media contents such as news articles, message boards, or even search volumes have gained popularity among researchers \citep{tet:07,antweiler2004all}. Whereas previous research considers low-frequency data, witnessing tremendous growth in the number of released news having shorter distances between the releases, it is tempting to ask a question if this almost ``continuous'' news feed can be used to measure a ``continuous'' changes in market sentiment. Even more interesting would then be to learn if such high-frequency news sentiment contains financially relevant information concerning stock price and volatility. To understand the evolution and dynamics of high-frequency news sentiment, we propose to use machine learning techniques to quantify sentiment from high-frequency news from the NASDAQ news platform, and we introduce a new high-frequency text-based sentiment measures to economists. Second, we propose to model the sentiment measures as a stochastic process and document how news sentiment influences volatility and price dynamics empirically by calibrating a sentiment-driven stochastic volatility model on data. Third, we provide explicit analytical moments that allow studying the non-trivial correlation of sentiment and volatility theoretically.

Sentiment and its role in financial markets have been attracting the attention of researchers from the early days. \cite{keynes1936general} argued that markets can fluctuate wildly due to investors' ``animal spirits,'' which can move prices in a way unrelated to fundamentals. Decades later, \citep{de1990noise} formalized the role of investor sentiment in financial markets arguing that uninformed noise traders basing their decisions on sentiment lead to noisy trading, mispricing, and excess volatility. The growing consensus that noise traders can impact stock markets in the short run is nicely surveyed by \cite{baker2006investor}.

Yet another decade later, the literature quantifying the investor sentiment and its impact on stock markets already grew into several strands. While sentiment is often measured using certain market-based variables as in \cite{baker2006investor}, market-based measures have the drawback of being equilibrium outcome of many economic forces other than investor sentiment \cite{da2014sum}. In turn, one can use survey-based measures including the University of Michigan Consumer Sentiment Index, or AAII investor sentiment survey, but these are available only in low frequencies. More recently, the textual analysis of media content became popular with this respect. Using daily content from a Wall Street Journal, \cite{tet:07} find that high media pessimism associated with low investor sentiment results in downward pressure on prices and upward pressure on volatility. \cite{antweiler2004all} study more than 1.5 million messages from message boards and find a significant effect on future volatility of the market. \cite{da2014sum} construct a sentiment index based on google search results and connect the sentiment with short-term return and increase in volatility.

It is worth noting that one should not comprehend news sentiment synonymously to investor sentiment. These two sentiments are related, most likely, in a Granger-cause relation \citep{tet:07} but not identical. News sentiment, a quantified news tone, may trigger the behavioral bias on agents' reactions. Nowadays, news articles, especially headlines, are written in a catchy way to gain readers' attention, and appeal to the readers' hungry for information. With such exciting news predominantly existing in the news platform nowadays, on one hand, investors intend to overreact, on the other hand, the following news gradually balance the biased news tone. As a consequence, the news sentiment, from over-pessimism/optimism, may revert to the level where it should stays. 

Another scenario is that news travels slowly \citep{hong2000bad}, the gradual-information-diffusion model of \citep{hong1999unified} fits this camp. Negative news, particularly, diffuses only gradually across investing public, leading to an underreaction to news.  Taken together, the news sentiment stochastics is akin to a mean-reverting process, resulting in the sentiment effect that yields a short-run departure between market price and its fundamentals. An understanding of investor behavior benefits from the analysis of the high-frequency news sentiment dynamics. 

We contribute to characterize news evolvement as a mean-reverting Ornstein-Uhlenbeck process and formalize it in a high-frequency setting. We introduce the news stochastics to economists in the asset pricing and mathematical finance studies. For instance, the decision making under uncertainty rests on the conditional information set that generates the probability distribution, as such, news enters the pricing mechanism exogenously and rules the volatility stochastics. The incomplete information renders a situation pointing to the decision making under uncertainty and risk. As pointed out by \cite{andersen2002empirical}, the classical continuous time models fail to account for the underlying dynamic evolution of stock price adequately. While a growing body of literature confirms the role of news and subsequent sentiment behavior on asset pricing and volatility, one possible innovation is to quantify the news process and incorporate it into the price and volatility process. 


Several insights can be gained from our analysis. First, high-frequency news sentiment can provide deeper empirical evidence on a well-documented short-lived effect of news and the consequent impact on asset prices and volatility. Second, we characterize a news life cycle from an over-optimistic/pessimistic tone reverting to a reasonable level as news is fully disclosed and analyzed eventually. It is useful to articulate the specifications that the model should be able to reflect reality.  Modeling the news evolution and calibrating it on high-frequency data, we evaluate the speed of news tone reversion in reality. 

Third, we verify the proposed framework by comparing the performance between the conventional stochastic volatility model \citep{hes:93} with the sentiment-driven stochastic volatility (SSV) model, in terms of shaping the actual evolution of asset price. Fourth, we present the theoretical properties of the proposed process and provide closed-form formulas for moments, co-moments and correlation. These efforts enable us to measure how sentiment impacts volatility theoretically and analytically. Finally, we end up the conclusions that the exogenous news sentiment stochastics impacts the threshold of volatility reversion, yielding a deferred reversion and prolonged volatility persistence. The instantaneous correlation between two separate Brownian motions driving the sentiment and volatility process is negative, akin to the leverage effect, in spite of insignificance.

We proceed the support vector machine (SVM) classifier to quantify the sentiment score for all \num{541750} news released in the \href{http://www.NASDAQ.com/news}{NASDAQ news platform} between January 3, 2012, to January 1, 2017. Further, we compute the aggregate sentiment score by averaging, at 15-min intervals, the sentiment scores of individual news published per 15-min period. The news in a high-frequency frame reveals a $U-$ shape news volume and also a $U-$ shape sentiment score, akin to a well-discovered a $U-$ shape trading volume. In terms of parameter calibration, we apply the nonparametric simulated maximum likelihood estimation by \cite{kristensen2012estimation} that constitutes an opportune estimation for the general class of the SSV model.

The paper is structured as follows. Section 2 details the high-frequency news,  the textual analytic tools and quantifies news sentiment. Section 3 introduces the sentiment-driven stochastic volatility model, and in section 4 we document the theoretical properties of the proposed process and provide closed-form formulas for moments, co-moments and correlation.  Section 5 describes simulated-based estimation, while section 6, using the estimation technique in section 5, undertakes the estimation for collected high-frequency sentiment and volatility of the S\&P 500. Section 7 concludes.


\section{High-frequency news sentiment and its quantification from text}

\subsection{Source of news}

While there are many possible sources of online news, most of the providers prohibit the application of automatic programs to download parts of their websites in the Terms of Service (TOS). Whereas the usage of web scrapers for noncommercial academic research is principally legal, these Terms of Service (TOS) are still binding \citep{truyens2014legal}. Even in the situation when the TOS is not a barrier, only limited message history may be available (such as the messages on Yahoo! Finance employed by \cite{antweiler2004all}). To ensure the sufficient news volume and frequency, we consider news articles that are available through the \href{http://www.NASDAQ.com/news}{NASDAQ news platform}.

In contrast to other sources, NASDAQ offers a platform for news and financial articles from selected contributors including leading media such as Reuters, MT Newswires, RTT news, or investment research firms such as Motley Fool, Zacks or GuraFocus. The news contents are classified as several categories e.g. stocks, economy, world news, politics, commodities, technology or fundamental analysis. News in the stocks category accounts for a big proportion where one finds the symbols assigned by NYSE, NASDAQ or other exchanges. The timestamp, the date, the contributor, the symbols, the title, and the complete text are all extracted using a self-written automatic web scraper and stored at our \href{https://sfb649.wiwi.hu-berlin.de/fedc/}{Research Data Center} of the IRTG 1792 at Humboldt-Universität zu Berlin. For more details, please refer to \cite{zha:wh:chen:bom:16}.

\begin{figure}[h]
    \centering
    \includegraphics[width=0.65 \textwidth]{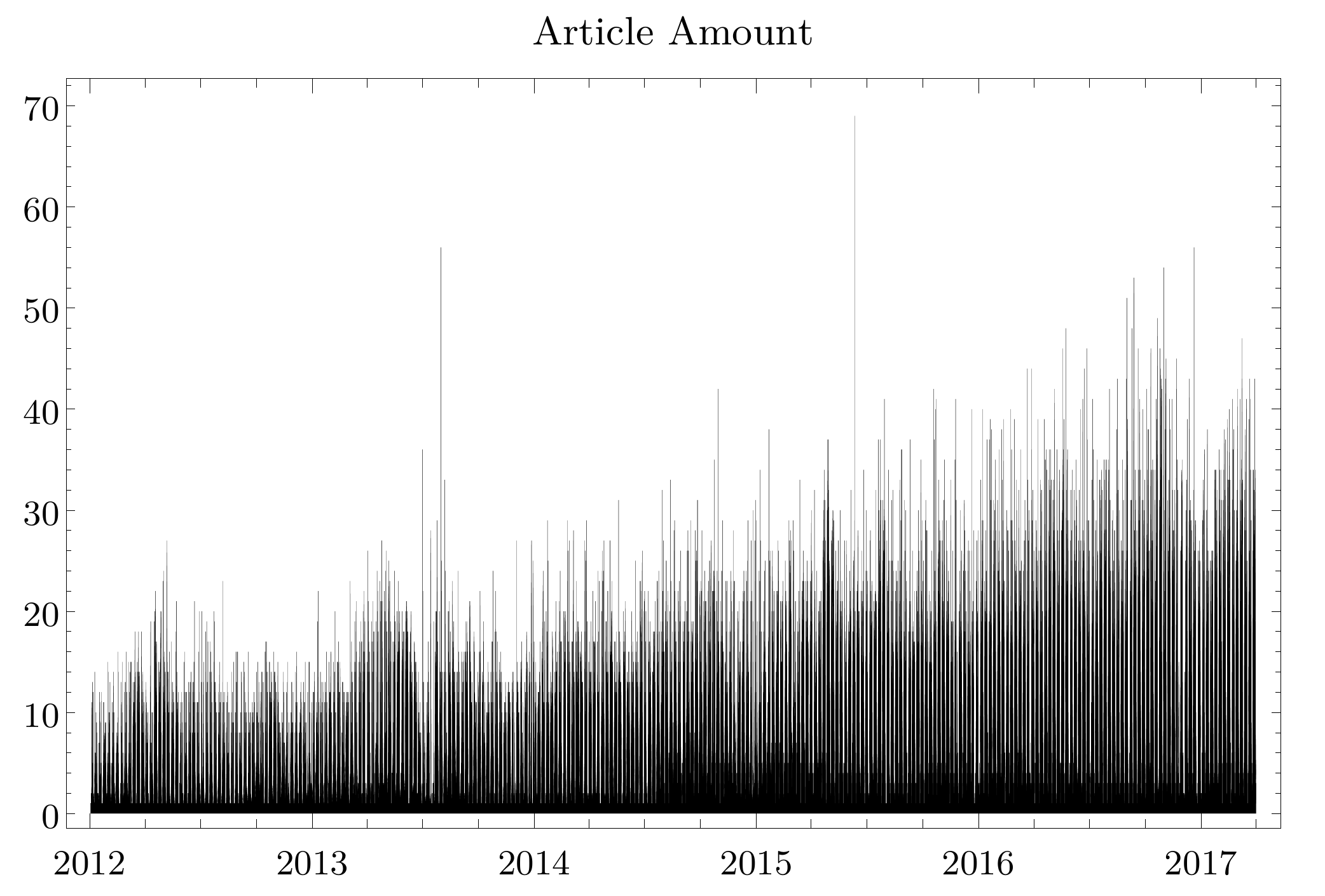}
    \caption{News arrival in 15-minute interval}
    \label{fig:artpday}
\end{figure}

In total, we collect \num{541750} articles during the period from January 3, 2012, to January 1, 2017. Figure \ref{fig:artpday} illustrates the volume of published articles per 15-min trading interval over time. We find that news posting is concentrated mainly during trading hours from 09:30 a.m until 04:00 p.m Eastern time, or equivalently working hours of reporters and financial analysts. One can observe an upward trend of news volume and the seasonality of publishing patterns. Intraday news usually peaks at market opening, declines at lunchtime and rises again before the market closes. The upward trend is the consequence of a growing number of contributors on the NASDAQ platform. While we collect around 10 to 20 published articles within the 15-minute interval during the years 2012, or 2013, the number of articles peak to 40, or 50 in 2016. 

\subsection{Textual analysis} \label{sec:textual}

The rapid development of natural language processing technologies (e.g. lemmatization, stemming, part-of-speech) has paved the way for automatic sentiment analysis. Note that the terms \textit{sentiment} and \textit{semantic orientation} are often used interchangeably in the textual analysis. Among the methods proposed for semantic orientation, the ``dictionary-based'' method stands for the most popular one such as the Harvard-IV dictionary or the \cite{loughran2011liability} lexicon. A particular effort made by \cite{loughran2011liability} is to create the lexicon for parsing financial articles. They extract all words occurring in at 5\% of \num{121217} 10-K reports, the further effort is to hire the experts who manually classify eligible words as positive, negative or neutral.

The overall semantic orientation of a sentence may differ from that of individual words. The focus has been switched from the lexicon-based method to the phrase/sentence structure model. Although lexicons help to detect the relevant vocabularies that carry better semantic comprehension, the overall sentiment of phrases is strongly dependent on the way that the phrases have been structured. One should not overlook the contextual cues embedded in the phrase structures.

The semantic orientation is also subject to the domain-specific use of language. A domain such as product reviews, the sentiment is often expressed with a combination of adjectives and adverbs, whereas financial sentiment tends to express the orientation by using expected favorable or unfavorable directions of events (e.g., the revenue is expected to increase). Given this fact, the economic or financial sentences essentially carry a strong semantic orientation but can be discounted while employing the knowledge from other domains.

\subsubsection{Support vector machine classifier}

Considering the complexity of sentiment distillation discussed previously, the support vector machine (SVM) approach is applied here for its ability of capturing the phrase/sentence structure information and adapting domain-specific use of language. Collecting financial and economic news texts as training dataset, we accumulate the domain-specific knowledge and use them to train the learning algorithm.

Given the regularized linear model, training data $(X_1, y_1), \ldots, (X_n, y_n)$ with $X_j \in \mathbb{R}^p$ and $y_j \in \{-1, 0, 1\}$ representing the sentiment classes, one calibrates the linear scoring function $s(X)= \omega^{\top} X + b $ via the regularized training error
\begin{equation}
n^{-1} \sum_{j=1}^n L\{y_j, s(X)\} + \lambda R(\omega)  \label{eq:regtrain}
\end{equation}
where $n$ is the number of sentences in the document, with $L(\cdot)$ as loss function, $R(\cdot)$ as regularization term and penalty $\lambda \geq 0$.  $X_j$ is the $j$th sentence that comprises a vector of words; $y_j$ is the corresponding semantic orientation, and is labeled by annotators as -1 (negative), 0 (neutral) or 1 (positive).

The first term in (\ref{eq:regtrain}) represents the training error that we aim to minimize, while the second one controls the complexity of set of models and is often named as regularizer. If a high capacity of set of function is employed, the resulting training error is smaller but the overfitting problem may emerge. Likewise, larger training errors tend to happen while a simpler regularizer is employed.

We have applied different loss functions. In terms of SVM, one may employ the Hinge loss
\begin{equation}
L\{y, s(X)\} = \max\{0, \, 1 - s(X) y\}
\end{equation}
or the Logistic likelihood $L(u) = \exp(-u)/\{1+\exp(-u)\}$. The least squares loss $L(u) = u^2 $ leads to the well known Ridge regression. As regularization term one may employ the $L_2$ norm
$R(\omega) = p^{-1} \, \sum_{j=1}^p \omega^2_j$ or the $L_1$ norm
$R(\omega) = \sum_{j=1}^p \, \lvert \omega_j \rvert$, given the calibration task a Lasso type twist.

\subsubsection{Financial phrase bank as training dataset}

The human-annotated financial phrase bank, constructed by \cite{Malo2014}, can be used for training and evaluating alternative models for financial and economic news texts. Having this training data, we are able to calibrate the scoring function $s(X)$ in (\ref{eq:regtrain}). The calibration approach is based on the stochastic gradient descent method for minimizing the loss function in (\ref{eq:regtrain}) that is written as a sum of differentiable functions.
The regularization parameter has been optimized by using 5 fold cross validation in which the data set is partitioned into 5 complementary subsets.  Four out of these 5 subsets are then combined to build the training data set.

In summary, we ran the predictive models and obtained best model accuracy in terms of this fitting technique by employing the hinge loss, the $L_1$ penalty and $\lambda = 0.0001$.  The mean accuracy for this training data set is 82\%, whereas the LM lexical projection, achieves only 64\% of accuracy.  A deeper analysis revealed that LM produced more often false negatives than the SVM method did. Concerning the accuracy of semantic orientation, we decide to stick to the sentiment quantified by the SVM method throughout this study.

Consequently, we obtain a huge vector $\hat \omega$ that enters the training function $s(X)$ which is now applied to the NASDAQ data set. Having the sentences ($Y_j$= -1, 0 or 1) classified by the SVM approach, we follow  \cite{antweiler2004all} to combine both negative and positive sentiment in the sentence level into an overall sentiment score in the document level:
\begin{equation}\label{eq:score}
B_{i,t} \; = \; \log(\;\!1 + n ^{-1}\sum_{j=1}^n\operatorname{\mathbf{I}}(\hat y_j=1)\;\!)\: - \: \log(\;\!1 + n ^{-1}\sum_{j=1}^n\operatorname{\mathbf{I}}(\hat y_j=-1)\;\!)
\end{equation}
where $n$ is the number of sentences in the document, $j$ is the index for sentence, $\hat y_j$ is the estimate of sentiment class by the SVM classifier. $B_{i, t}$ is the sentiment score of $i$-th news articles on the intraday interval $t$. One can easily observe that $B_{i, t} < 0$ holds if the polarity of the text is negative, while $B_{i, t} = 0$ indicates neutrality and $B_{i, t} > 0$ suggests a positive polarity. As a result, we obtain $B_{i, t}$ for each document and each intraday interval of our sample period. We then average sentiment scores $B_{i, t}$ within the $t$-th interval to yield the $B_t$.

\subsubsection{Intraday sentiment scores}
We proceed the SVM classifier to quantify the sentiment score defined in (\ref{eq:score}) with the values bounded between -1 and +1 for all \num{541750} news released in the \href{http://www.NASDAQ.com/news}{NASDAQ news platform} between January 3, 2012 to January 1, 2017. Further, we compute the aggregate sentiment score by averaging, at 15-min intervals, the sentiment scores of individual news published per 15-min period. Given 15-min intervals per trading day, we obtain 26 sentiment scores generated from 09:30 a.m until 04:00 p.m Eastern time during every trading day.

\begin{figure}[ht!]
\begin{center}
\begin{tabular}{cc}
\includegraphics[width=0.5 \textwidth]{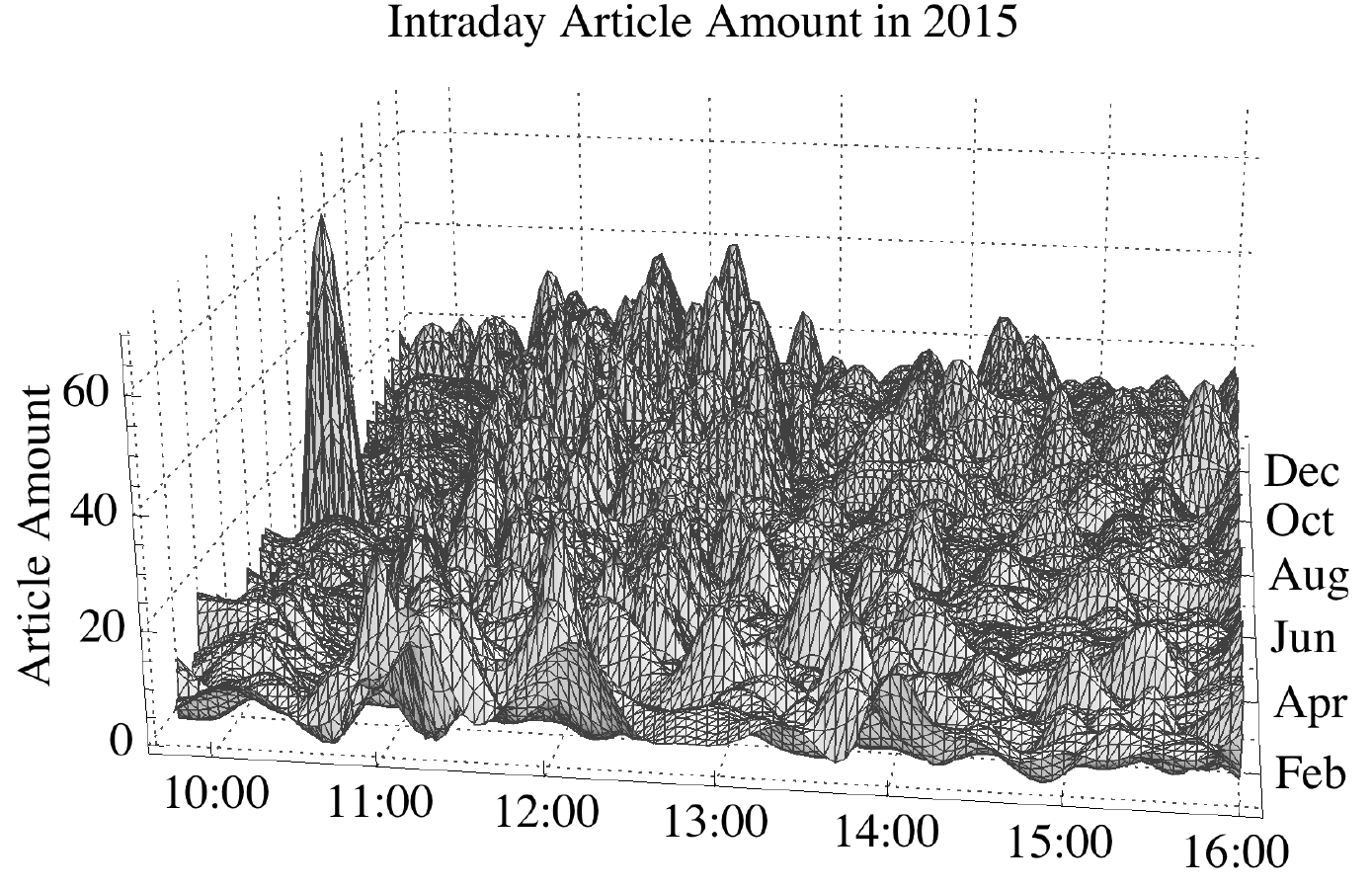} &
\includegraphics[width=0.5 \textwidth]{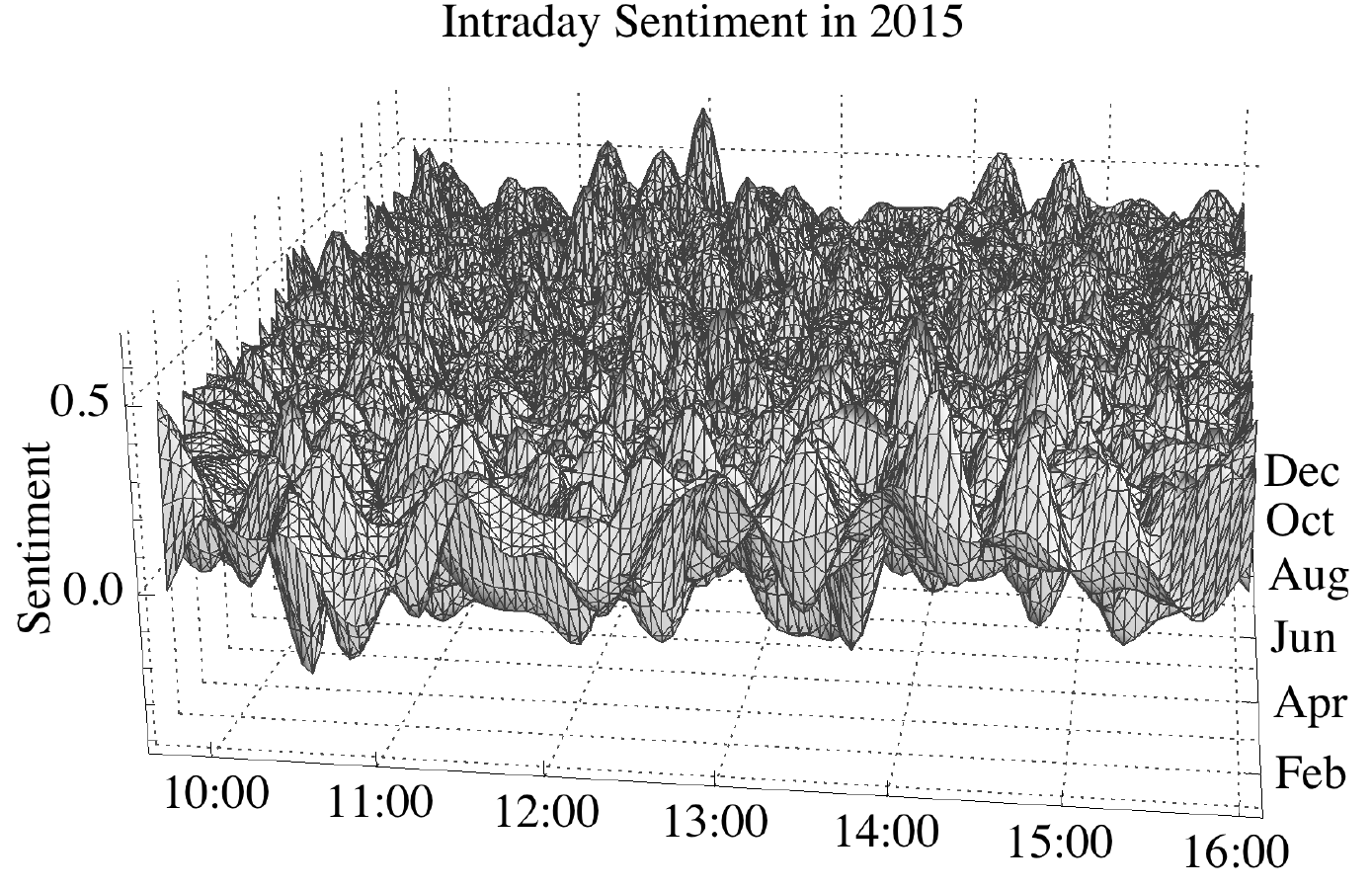}  \\
\end{tabular}
\vspace{-0.5cm}
\caption{Intraday news amount and sentiment surface}\label{fig:Intraday}
\end{center}
\end{figure}

Taking news published in 2015 as an illustrative example, we depict the article volume across trading hours and over time and show it in the left panel of Figure \ref{fig:Intraday}. Numerous amount of news are arriving after the market is open. The amounts of published articles is increasing until the lunch time, whereas after the lunch  the quantities decline with slight rebound before the close.

Interestingly, the general tone of news depicted in the right panel of Figure \ref{fig:Intraday} is rather positive at the market open. The news tones captured by sentiment scores vary over trading time, and again turn optimistic at the market close. It seems that the investment research firms and news media tend to release the news with positive tone at the market open, which may be interpreted as a kind of strategy used to motivate trading in the buying side. Or the reporters are just simply optimistic in the beginning of day. The positive tones observed at the market close convey the similar signal. In this way, they comfort investors to keep holding their positions overnight. Another explanation for the $U-$ shape alike sentiment score over trading hours is that the positive news basically are majority and account for almost 70\%, resulting in a ``visible'' optimistic mood at the market open and close.

\begin{figure}[ht!]
\begin{center}
\includegraphics[width=0.6 \textwidth]{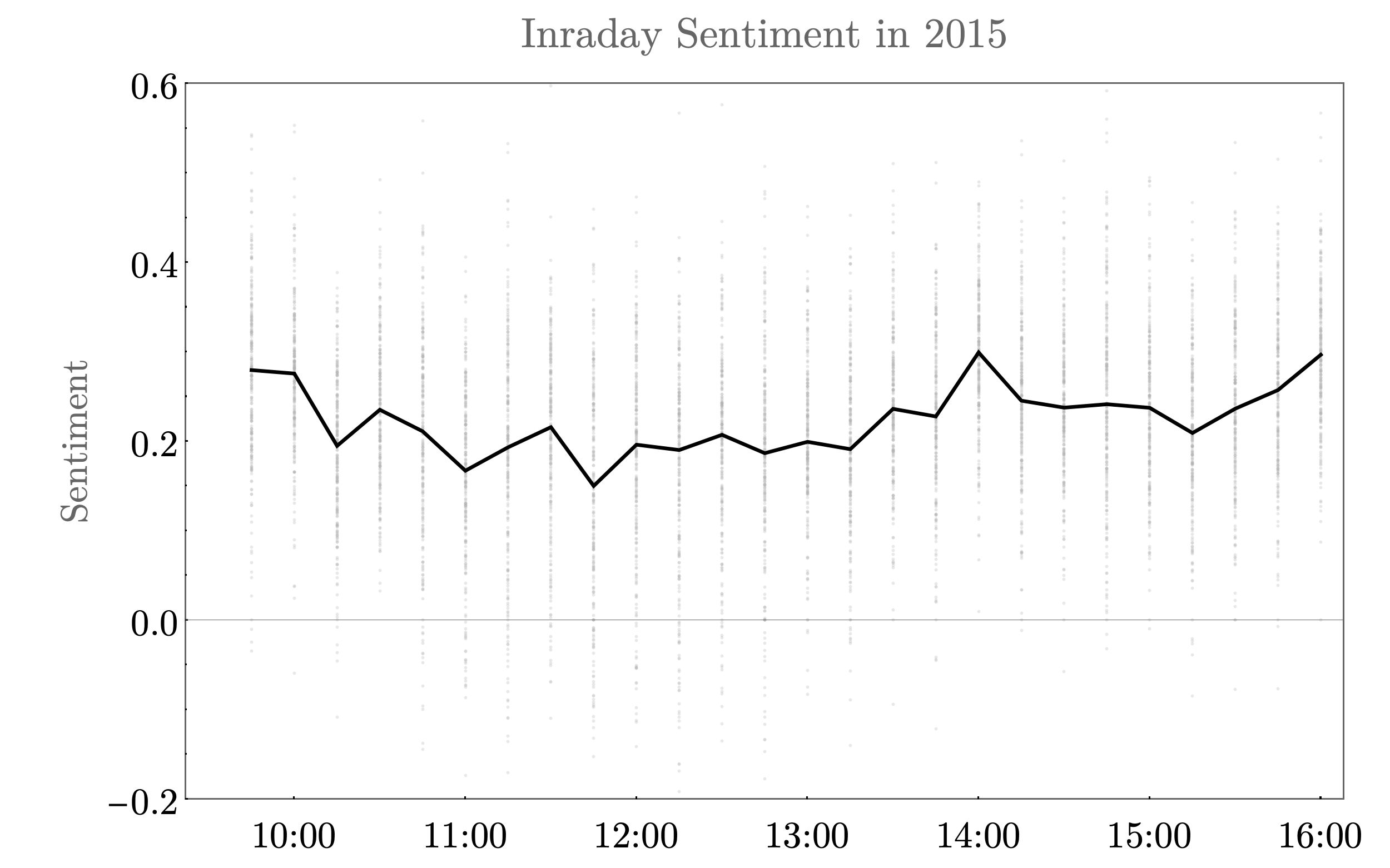}\vspace{-0.5cm}
\caption{Intraday news amount and sentiment surface}\label{fig:Intraday2}
\end{center}
\end{figure}

Figure \ref{fig:Intraday2} complements the previous three dimensional plot with the mean value of sentiment through the trading hours confirming the $U-$ shape.

%
%

\section{Sentiment-driven stochastic volatility model} \label{sec:SSV} 
\subsection{Stochastic sentiment process }
News arrives randomly and its quantified content, news sentiment, may behave like a Brownian motion process.  However, exaggerated headlines become popular nowadays due to media competition. Given the vast array of media now available in the traditional way or online, to compete for attention, reporters often resort to sensationalist headlines to target a different audience segment.  Especially, once the reporters hold the first-hand news,  intriguing headlines and probably an overstated news content from them may substantially emerge. The following news may balance the overstated ones as more information arrives. 

Another theme is that news, especially bad news, travels slowly \citep{hong2000bad}, as the consequence of sequential information arrival, scarce available information or strategically mitigated overreaction.  As media compete with each other, news sentiment will correct itself until it reaches a relatively reasonable level. Based on this mechanism, we postulate news sentiment behaving like a mean-reverting (Ornstein-Uhlenbeck) stochastic process as:
\begin{equation}\label{eq:sent}
    dS_t = \lambda_S (\mu_S - S_t) dt + \sigma_S dW_{s,t}
\end{equation}
where $\mu_S$ is the long-run mean level of news sentiment, $\lambda_S$ governs the speed of mean-reversion, and $dW_{s,t}$ is the random innovation term with the diffusion parameter, $\sigma_S$. The smaller the value $\lambda_S$ is, the linger the excursions from mean last. As a consequence, news sentiment (hereafter being abbreviated to sentiment) may exhibit some predictable periodicities. (\ref{eq:sent}) implies the increment in a random sentiment can be composed of two parts, the predictable part in the drift term and the unpredictable part in the diffusion term. 
Here $S_t$ is quantified by the textual analysis introduced in section \ref{sec:textual} and news are grouped in a 15-min interval.

\subsection{Joint evolution of sentiment, price and volatility process}
The overstated or understated news sentiment may trigger investors' animal spirit, resulting in a subsequent impact on the price and volatility dynamics \citep{tet:07}. As pointed out by \cite{andersen2002empirical}, the classical continuous time models fail to account for the underlying dynamic evolution of stock price adequately. While a growing body of literature confirms the role of news and subsequent sentiment behavior on asset pricing and volatility, one of the possible improvements is to quantify the news process and incorporate it into the price and volatility process. With these aims, we contribute to a close-to-reality specification.

So far, in the empirical investigation, the sentiment effects on return and volatility are explored separately and haven't yet been formulated in a unified framework in the literature. The conventional continuous-time equity return model emphasizes the role of volatility in term of ``risk-return tradeoff hypothesis'' on the price process. The volatility, however, is latent and requires selecting the proxies \citep{andersen2002empirical}. Given that news and sentiment drive the financial markets presented in the current literature (\cite{zha:wh:chen:bom:16}), we aim to propose a continuous-time framework in which the sentiment will drive volatility and returns.

Let $p_t$ denote the time-$t$ logarithmic asset price, we propose a \textit{sentiment-driven stochastic volatility} (SSV) model and postulate the following dynamics of the instantaneous returns, volatility and sentiment:
\begin{align}\label{1}
    \begin{split}
    dS_t &= \lambda_s (\mu_s - S_t) dt + \sigma_s dW_{s,t} \\
    dP_t &= (\mu_p - \exp(V_t)/2 ) dt +  \exp(V_t/2) dW_{p,t} \\
    dV_t &= (\mu_v + \beta_v (S_t - \mu_s)^2 - \gamma_v V_t) dt + \sigma_v dW_{v,t}
    \end{split}
\end{align}
with $corr(dW_{p,t},dW_{v,t})=\rho_{pv}$, and $corr(dW_{s,t},dW_{v,t})=\rho_{sv}$. Empirically, the correlation between the two separate Brownian motions, $\rho_{p,v}$, is negative, which is so called continuous time leverage effect. Building on a specification used by \cite{andersen2002empirical}, we further link the mean adjusted sentiment to volatility via parameter $\beta_v$. In addition, sentiment process enters the processes of price and volatility simultaneously, and predicts the dynamics of instantaneous return and volatility. $\gamma_v$ defines the speed of reverting to the mean level, while the threshold of reverting to the mean level is influenced by sentiment. With the presence of sentiment either positive or negative value, volatility only gets reverted as it exceeds the sum of long-run mean level, $\mu_v$, and the sentiment-related component $\beta_v (S_t - \mu_s)^2$ if the $\beta_v$ coefficient is significantly positive. It may capture the extreme volatility caused by a rising breaking point of reversion stemmed from the sentiment component. Also, the speed of reverting to the mean level could be subject to the presence of sentiment. It can be understood that volatility may behave very persistent as news/sentiment are consecutively presenting in the markets, which can be empirically documented by \cite{antweiler2004all}. 

\section{Sentiment and Volatility: Moments, Co-moments, and Correlation}

In this section, we study the theoretical properties of the proposed process and provide closed-form formulas for moments and co-moments of the sentiment and volatility. These insights are important for further applications as it enables us to measure how sentiment impacts volatility theoretically and analytically. 

The general computational strategy is to 
determine $\E[S_tV_t], E[S_t], \E[S^2_t], \E[V_t], \E[V^2_t]$ 
from the proposed stochastic evolution of the processes. This results in the formulas that may seem complicated, but it exists in the closed form. We show, under some conditions, the moment properties of the process will degenerate to the variance process of the SV model. We collect the results in the following theorems and properties and relegate proofs to the Appendix \ref{sec:app}.

Before introducing the theorems adhered to the proposed model, we briefly describe the details of how we use the necessary propositions as a means to obtain the desired theorems. As the proof of the theorems will require computation of the moments of the processes $S_t$ and $V_t$, we articulate the Propositions \ref{S} -- \ref{EV2}. We start with the moments of the sentiment process $S_t$.


\begin{proposition} \label{S}
The first two moments of the process $S_t$ are given by
\begin{equation}
\E[S_t] =   e^{-\lambda_s t} S_0 + \mu_s[1 - e^{-\lambda_s t}],
\end{equation}
and
\begin{equation}
\E[S^2_t] =   \frac{\sigma^2_s}{2 \lambda_s} \cdot [1 - e^{-2\lambda_s t}] + (\E[S_t])^2.
\end{equation}
\end{proposition}

As the computation of expectations involving the process $V_t$ requires also knowledge of the third and 
the fourth moment of $S_t$, let us list them here.
\begin{equation}
 \E[S^3_t] = \E[S_t]\cdot \left((\E[S_t])^2 + 3 \cdot (\E[S^2_t] - (\E[S_t])^2) \right)
\end{equation}
\begin{equation}
 \E[S^4_t] = (\E[S_t])^4 + 6 \cdot (\E[S_t])^2 \cdot (\E[S^2_t] - (\E[S_t])^2) + 3 \cdot (\E[S^2_t] - (\E[S_t])^2)^2
\end{equation}
This follows from the normality of the process $S_t$. Having defined the moments of $S_t$, we describe the first and the second moment of the process $V_t$.

\begin{proposition} \label{V}
The first moment of $V_t$ is given by
\begin{equation}
 \E[V_t] = e^{-\gamma_v t} \cdot V_0 +  \int_0^t e^{\gamma_v (u-t)} \left[ 
\mu_v + \beta_v \left(\E[S^2_u] - 2 \mu_s \E[S_u] + \mu_s^2 \right) 
 \right] du.
\end{equation}
\end{proposition}

Note that in the case of $\beta_v=0$, we have the mean level of variance process that is degenerated to the case in \cite{hes:93}. Non-zero positive $\beta_v>0$ increases mean of volatility. Hence this additionally introduced sentiment process increases volatility, and higher volatility of turbulent markets could hence be explained by large amounts of exaggerated news. Propositions \ref{SV} and \ref{EV2} further describe the relation of $S_t$ and $V_t$ processes as well as second moment of $V_t$.

\begin{proposition} \label{SV}
The expectation  $\E[S_t  V_t]$ is given by
\begin{multline}
\E[S_t  V_t] =  e^{-(\gamma_v + \lambda_s)t}  \cdot S_0 V_0 \\ +  \int_0^t  
e^{(\gamma_v + \lambda_s)(u-t)}
\Biggl(
\lambda_s\mu_s \E[V_u] + \sigma_s\sigma_v \rho_{sv}
 +  [\mu_v + \beta_v \mu_s^2] E[S_u] \\
 - 2   [\beta_v \mu_s]  E[S^2_u] 
 +    \beta_v \E[S^3_u] \Biggr) du.
\end{multline} 
\end{proposition}

\begin{proposition} \label{EV2}
The second moment of $V_t$ is given by
\begin{multline}
 \E[V^2_t] = e^{-2 \gamma_v t} \cdot V_0^2 +  \int_0^ t e^{2 \gamma_v (u-t)}
 \Biggl( 2 \left[\mu_v + \beta_v \mu_s^2\right] \E[V_u] - 4 \beta_v \mu_s \E[S_uV_u] \\
 + 2 \beta_v \E[S_u^2 V_u] + \sigma_v^2
 \Biggr) du.
\end{multline}
\end{proposition}

The impact of sentiment on these quantities is non-trivial and depends on the rest of the parameters. However, by fine-tuning a particular parameter and remaining others, we may be able to make further influences on these propositions. In proposition \ref{SV}, assuming that all other parameters are fixed, a higher $\rho_{vs}$ will inflate $\E[S_t  V_t]$ in (11) and also the resulting correlation $\rho(S_t, V_t)$ in (19). Recall that $\rho_{vs}$ measures the instantaneous correlation between two separate Brownian motions driving the sentiment and volatility process. The sign of the $\E[S_t  V_t]$  and their correlation documented in theorem \ref{theorem:rho} hinge on the value of $\rho_{vs}$. Likewise, in proposition $\ref{V}$, the value of $\E[V_t]$ is governed by $\beta_v$. In terms of $\beta_v$'s presence, the threshold of reverting to the mean volatility level is elevated, leading to a higher mean level. 

Note that Proposition \ref{EV2} requires knowledge of $\E[S_t^2 V_t]$ determined in the following lemma.

\begin{lemma} \label{ES2V} The expectation of $ \E[S_t^2 V_t]$ is given by
\begin{multline}
 \E[S_t^2 V_t] = e^{-(\gamma_v  + 2 \lambda_s) t} \cdot S_0^2 V_0 + \int_0^t e^{(\gamma_v  + 2 \lambda_s)(u- t)}
 \Biggl( 2 \sigma_v\sigma_s \rho_{sv} \E[S_u] + [\mu_v + \beta_v \mu_s^2] \E[S_u^2] \\ 
 -2 \mu_s\beta_v \E[S_u^3] + \beta_v \E[S_u^4] + 2 \lambda_s \mu_s \E[S_u V_u] + \sigma_s^2 \E[V_u] \Biggr) du.
\end{multline} 
\end{lemma}

\begin{theorem}
\label{th1}
Expectations $\E[S_t]$, $\E[V_t]$, variances $\text{\rm Var}(S_t)$, $\text{\rm Var}(V_t)$ and the covariance
$\text{\rm Cov}(S_t,V_t)$ are given by the following formulas:
{
\small
\begin{equation}
\E[S_t] = 
e^{-\lambda_s t} \left(S_0+\mu_s \left(e^{\lambda_s t}-1\right)\right)
\end{equation}

\begin{dmath}
\E[V_t] =  \frac{e^{-(\gamma_v+2 \lambda_s)t}}{2 \gamma_v \lambda_s (\gamma_v-2 \lambda_s)} \cdot
\Biggl[
\beta_v \gamma_v e^{\gamma_v t} \left(2 \lambda_s \mu_s^2+2 \lambda_s S_0^2-4 \lambda_s \mu_s S_0-\sigma_s^2\right)-2 \lambda_s e^{2 \lambda_s t}
   \left(\beta_v \gamma_v \mu_s^2-\beta_v \sigma_s^2+\gamma_v \mu_v-2 \lambda_s \mu_v+\beta_v \gamma_v S_0^2-2 \beta_v \gamma_v
   \mu_s S_0-\gamma_v V_0 (\gamma_v-2 \lambda_s)\right)+(\gamma_v-2 \lambda_s) e^{(\gamma_v+2 \lambda_s)t} \left(\beta_v \sigma_s^2+2 \lambda_s
   \mu_v\right)\Biggr]
\end{dmath}

\begin{equation}
\text{\rm Var}(S_t) = 
\frac{\sigma_s^2 e^{-2 \lambda_s t} \left(e^{2 \lambda_s t}-1\right)}{2 \lambda_s}
\end{equation}

\begin{dmath}
\text{\rm Var}(V_t) = \frac{e^{-2 (\gamma_v+2 \lambda_s)t}}{2 \gamma_v \lambda_s^2 (\gamma_v-2 \lambda_s)^2
   (\gamma_v-\lambda_s) (2 \gamma_v-\lambda_s) (\gamma_v+\lambda_s) (\gamma_v+2 \lambda_s)} \cdot \Biggl[-a+b-c-d+e+f\Biggr]
\end{dmath}

\begin{dmath}
\text{\rm Cov}(S_t,V_t)=
\frac{\sigma_s e^{- (\gamma_v+3 \lambda_s)t} }{\gamma_v \lambda_s (\gamma_v-2 \lambda_s) (\gamma_v+\lambda_s)}
\cdot
\Biggl[\beta_v \sigma_s \left(\gamma_v^2-\gamma_v \lambda_s-2 \lambda_s^2\right) (S_0-\mu_s) e^{(\gamma_v+2
   \lambda_s) t}+\lambda_s e^{2 \lambda_s t} (-2 \beta_v \mu_s \sigma_s (\gamma_v+\lambda_s)-\gamma_v \rho_{sv} 
   \sigma_v (\gamma_v-2 \lambda_s)+2 \beta_v
   S_0 \sigma_s (\gamma_v+\lambda_s))-\beta_v \gamma_v \sigma_s (\gamma_v+\lambda_s) (S_0-\mu_s) e^{\gamma_v t}
   +\gamma_v \lambda_s \rho_{sv} \sigma_v (\gamma_v-2 \lambda_s) e^{(\gamma_v+3 \lambda_s)t}\Biggr]
\end{dmath}

where $a$, $b$, $c$, $d$, $e$, and $f$ are given in Appendix (\ref{sec:app}).
}
\end{theorem}

Finally, having derived moments and co-moments of the process $S_t$ and $V_t$, we may compute their correlation.

\begin{theorem}\label{theorem:rho}
 The correlation coefficient $\rho(S_t, V_t)$ is given by 
 \begin{equation}
\rho(S_t, V_t) = \frac{\E[S_t  V_t] - \E[S_t] \cdot \E[V_t]}{\sqrt{ \E[S^2_t] - (\E[S_t])^2} 
\cdot \sqrt{ \E[V^2_t] - (\E[V_t])^2}}, 
 \end{equation}
 where the expectations entering this formula are determined in Propositions \ref{S} -- \ref{EV2}.
\end{theorem}

Note that all integrals in this section admit an analytical form due to the fact that all integrands are simple exponentials. Note that the impact of sentiment on the correlation between sentiment and volatility is non-trivial, and depends on chosen parameters. 

Since the analytical forms derived above are complicated and do not show the impact of sentiment directly, we would like to illustrate it with parameters fixed to those estimated later on real data. 
\begin{figure}[ht!]
\begin{center}
\includegraphics[width=0.6\textwidth]{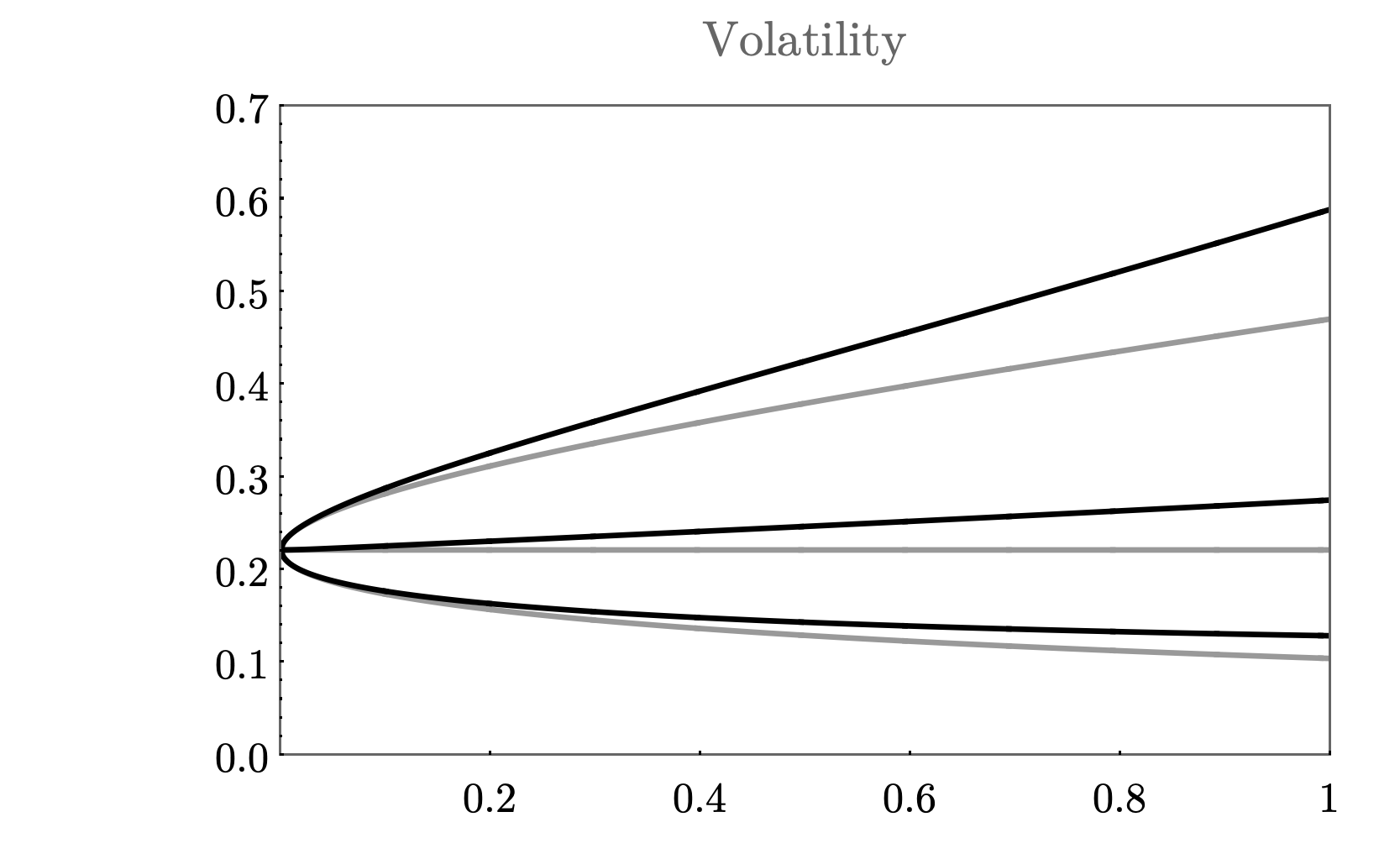}  \\
\caption{Impact of sentiment on the volatility. First two moments of the volatility process from (\ref{1}) with $\beta_v=0$ are shown in gray, and $\beta_v>0$ in black.}
\label{fig:moments}
\end{center}
\end{figure}
Figure \ref{fig:moments} illustrates impact of sentiment on the volatility. We compare the two processes with identical parameters except $\beta_v$ which is fixed to zero for illustration. Figure \ref{fig:moments} documents that positive beta increases volatility of the price process as well as its variance (so called volatility of volatility).

\section{Simulated-based estimation on empirical data}

Estimation of sentiment-driven stochastic volatility model is challenging due to the unknown closed-form representation of the likelihood function, and a more general computational framework for empirical validation of the SSV model is demanded. We, therefore, opt for the nonparametric simulated maximum likelihood estimation (NPSMLE) by \cite{kristensen2012estimation}  to recover parameters consistently and efficiently in similar situations. In this section, we discretize the framework of the SSV model presented in (\ref{sec:SSV}), and apply the NPSMLE that constitutes an opportune estimation for the general class of SSV model.

\subsection{NPSMLE}
Assuming the processes $(y_t,x_t), y_t\in \mathbb{R}, x_t \in \chi_t$, we desire the conditional density $p_t(y|x;\theta)$ based on the parametric model between them with $\theta$ as the parameter set:
\begin{equation}\label{eq:parEq}
      y_t=q_t(x_t,\varepsilon_t;\theta), \hspace{3mm} t=1,\ldots,T
\end{equation}
where $\varepsilon_t$ is an i.i.d sequence with known distribution $F_\varepsilon$. $x_t$ usually are exogenous explanatory variables or lagged dependent variables $y_t$. The density of $y_t$ conditional on several exogenous or lagged dependent variables may not guarantee a closed-form representation or itself can even be unknown similar to our case.

\cite{kristensen2012estimation} propose a general way of estimation via approximating the conditional density through simulations. The simulated version of conditional density $p_t(y|x;\theta)$ can be achieved by generating $N \in i.i.d$ draws from the predefined models, it therefore yields $y_{t,i}^\theta$
 where $i=1,\ldots,N$. These $N$ simulated $i.i.d$ random variables, $\{y_{t,i}^\theta\}_{i=1}^N$, follow the target distribution: $y_{t,i}^\theta \sim p_t(y|x;\theta)$. Hence, they in turn can be used to nonparasitically estimate via kernel methods such as
\begin{equation}\label{eq:kernel}
      \hat{p}_t(y_t|x_t;\theta)=\frac{1}{N}\sum_{i=1}^{N}K_h(y_{t,i}^\theta-y_t)
\end{equation}
where $K_h(\upsilon)=K(\upsilon)/h^k, K: \mathbb{R}^k \mapsto \mathbb{R}$ is a generic kernel, and $h>0$ a bandwidth. Under regularity conditions on $p_t$ and $K$, \cite{kristensen2012estimation} show :
\begin{equation}\label{eq:Asykernel}
      \hat{p_t}(y_t|x_t;\theta) = p_t(y_t|x_t;\theta)+ \mathcal{O}_p(1/\sqrt{Nh^k})+\mathcal{O}_p(h^2), \hspace{6mm} N \rightarrow \infty
\end{equation}
The application of the NPSMLE technique to the Cox-Ingersoll-Ross model with a known transition density and true parameters confirms that as the number of simulated observation increases, the biases, variance, and root mean squared errors (RMSEs) decrease dramatically, see Table 1 of \cite{kristensen2012estimation}. Bias due to the first-order approximation used in the construction of density and the variance resulting from the Monte Carlo integration are two sources of the discretization approximation error for true density. Besides, the NPSMLE appears to be quite robust towards the choice of bandwidth.

Having the convergence to probability asymptotically shown in (\ref{eq:Asykernel}), we can construct the simulated MLE based on (\ref{eq:kernel}):
\begin{equation}\label{simMLE}
      \hat{\theta}= \arg \max_{\theta \in \Theta} \hat{L}_T(\theta),  \hspace{8mm} \hat{L}_T(\theta)= \sum_{t=1}^{T}\mbox{log}\hat{p}_t(y_t|x_t;\theta)
\end{equation}
It's suggested to use the same draws for all values of $\theta$, and the same batch of draws from $F_\varepsilon(\cdot), \{\varepsilon_i\}_i^N$ in (\ref{eq:parEq}), across $t$.

Under the regularity conditions, (\ref{simMLE}) implies that $\hat{L}_T(\theta)\overset{p}{\to} L_T(\theta) $ as $N \rightarrow \infty$ for a given $T \geq 1$. The simulated MLE estimator, $\hat{\theta}$, retains the sample properties as the infeasible MLE estimator as $T, N \rightarrow \infty$ under suitable conditions.

\subsection{Estimation on high frequency data}
We implement the proposed estimation to the intraday news scrapped from \href{http://www.NASDAQ.com/news}{NASDAQ news platform}. The dramatically increasing news volume in the NASDAQ platform since 2015 ideally represents the continuous news feeds and therefore supports the proposed sentiment stochastic process in the continuous time framework. We start with the discretization of continuous processes in (\ref{eq:sent}) and (\ref{1}) and estimate them in practice via NPSMLE approach.

A diffusion process can be approximated by various discretization schemes up to a given level of precision. Furthermore, discretization schemes \citep{kloeden1992numerical,bruti2007approximation} can be used to simulate observations from the model for any given level of accuracy, enabling NPSMLE.

The first aim is to discretize the stochastic sentiment process in (\ref{eq:sent}) and obtain the estimates $\theta=(\lambda_s, \mu_s, \sigma_s)$. We collect 6500 discrete time 15-minute observations with the $t-s=1/(250\times26)$ between observations covering 250 trading days during the period of January 1, 2015 -- December 31, 2015. The period is of interest because of the Chinese stock market turbulence, spreading fear worldwide.  A three-week plunge has knocked about 30\% off Chinese shares since mid-June 2015. It reflects that investors have been piling in, encouraged by falling borrowing costs as the central bank loosened monetary policy. Every trading day, we consider 26 observations from 09:30 to 16:00 EST. Given a set of parameter values in the parameter space and initial observation $y_s$, we simulate paths using the Euler scheme. The time interval $t-s$ is further divided into $M$ subintervals, and we recursively compute for $m=0,\ldots,M-1$:
\begin{equation}\label{eq:discre}
    u_{m+1}^i = u_{m}^i+\lambda_s(\mu_s - u_{m}^i)\delta + \sigma_s \delta^{1/2} W_{m+1}^i,, \hspace{3mm}  m=0,\ldots,M-1
\end{equation}
where $u_0^i=y_s$, $\delta=\frac{t-s}{M}$, $W_{m+1}^i\sim$ N(0,1), and let $\hat{y}_{t,i}^\theta = u_M^i$ for (\ref{eq:kernel}), the $i$th simulated observation of $y_t$ conditional on $y_s$. Having the generated $\{y_{t,i}^\theta\}_{i=1}^N$, we are able to nonparametrically estimate via the kernel methods for the density of the simulated observations shown in (\ref{eq:kernel}) where a Gaussian kernel is applied and the bandwidth, $h$, is decided by Silverman's rule of thumb. The likelihood function based on the estimated density can be derived and can be maximized. The parameter set $\theta$ is therefore calibrated given the maximum value of log-likelihood function.

Analogously, the discretization for the SSV model in (\ref{1}) can be implemented in the same way, yielding the resulting discretized formulas as:
\begin{align}\label{eq:DisModel2}
 \begin{split}
    u_{s,m+1}^i &= u_{s,m}^i+\lambda_s (\mu_s - u_{s,m}^i)\delta + \sigma_s \delta^{1/2} W_{s,m+1}^i   \\
    u_{p,m+1}^i &= u_{p,m}^i +(\mu_p - \exp(u_{v,m}^i)/2)\delta + \exp(u_{v,m}^i/2) \delta^{1/2} W_{pv,m+1}^i \\
    u_{v,m+1}^i &= u_{v,m}^i+(\mu_v + \beta_v (u_{s,m}^i - \mu_s)^2 - \gamma_v u_{v,m}^i)\delta + \sigma_v \delta^{1/2} W_{sv,m+1}^i
 \end{split}
\end{align}
where $m=0,...,M-1$. The parameter vector to be estimated is $\theta=(\lambda_s, \mu_s,\sigma_s, \mu_p, \mu_v, \gamma_v, \beta_v, \sigma_v, \rho_{pv}, \rho_{sv})^{'} \in \mathbb{R}^{10} $. The first three parameters characterize the sentiment process, $\mu_p$ governs the mean of the log price process, and $\gamma_v, \mu_v, \beta_v, \sigma_v$ are decisive for the evolution of volatility process. The additional markup in reverting-threshold is attributed to $\beta_v(S_t-\mu_s)^2$.

The interplay among three Brownian motions is modeled as $W_{pv,m+1}^i = \sqrt{1-\rho_{pv}}W_{p,m+1}^i + \rho_{pv} W_{v,m+1}^i$, and $W_{sv,m+1}^i = \sqrt{1-\rho_{sv}}W_{s,m+1}^i + \rho_{sv} W_{v,m+1}^i$, where $W_{p,m}^i, W_{v,m}^i$, and $ W_{s,m}^i$ are i.i.d standard normal random variables with the correlation $\rho_{pv}$ between the first two Brownian motions and $\rho_{vs}$ between the last two Brownian motions.

\section{High frequency sentiment and volatility of the S\&P 500}

For the estimation, we assume that all processes are observable, and we search for the data that enables us to sketch high-frequency news sentiment, prices, as well as volatility. Here we follow \cite{ait:kim:07} who infers volatility from derivative prices. While we quantify the high frequency sentiment from news articles, data about prices and volatility are obtained from the Tick Data, Inc.\footnote{http://www.tickdata.com}. We use 15-minutes S\&P 500 futures contracts traded on the Chicago Mercantile Exchange (CME), and for the volatility, we use the VIX futures traded on the CBOE Futures Exchange. We collect the data for the same period as sentiment, that is January 1, 2015 -- December 31, 2015.

\begin{figure}[ht!]
\begin{center}
\includegraphics[width=\textwidth]{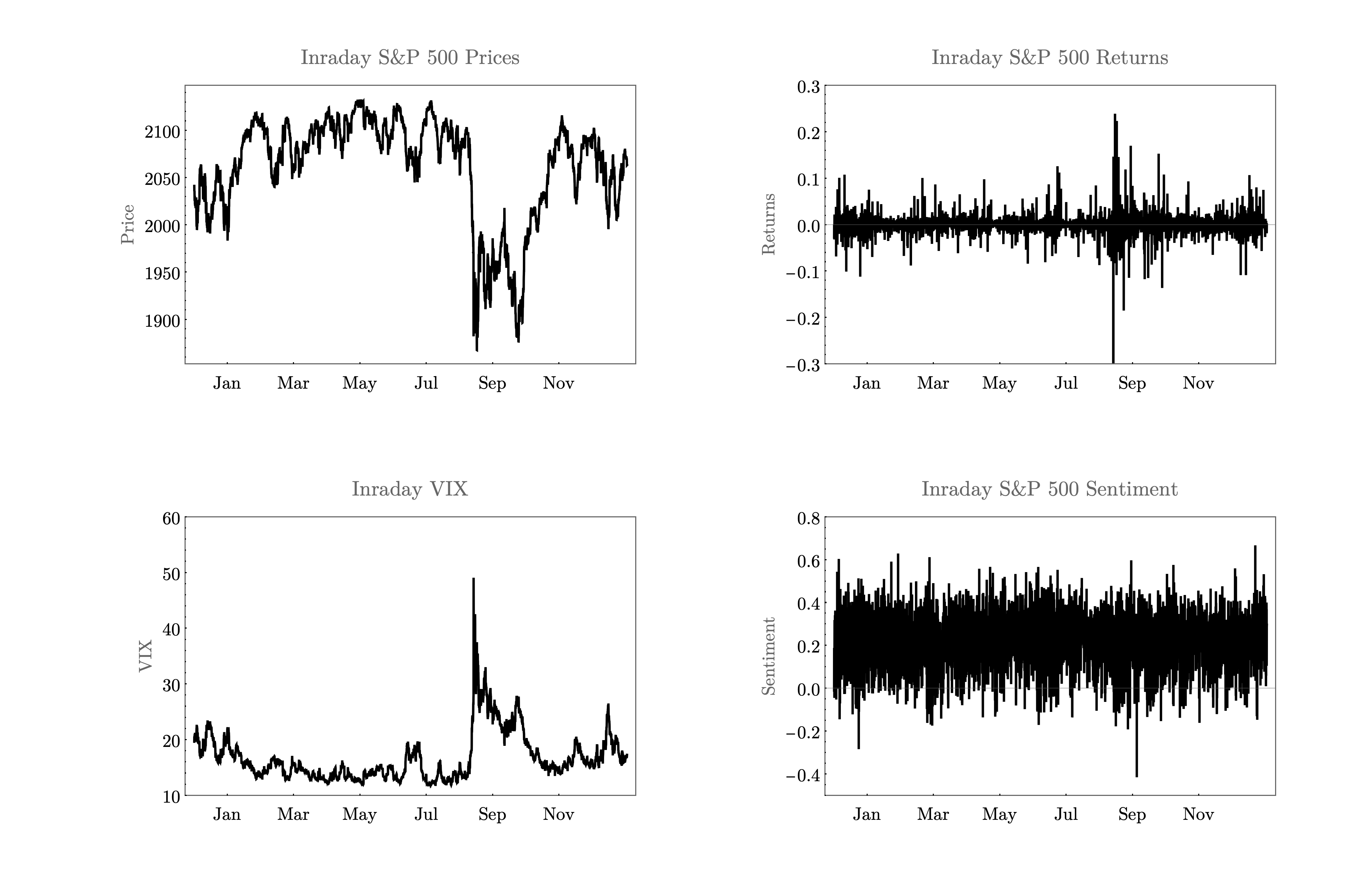}  \\
\caption{Intraday S\&P 500 prices, returns, VIX, and sentiment data spanning the period from January 1, 2015 until December 31, 2015.}
\label{fig:data}
\end{center}
\end{figure}

Figure \ref{fig:data} plots the intraday prices, returns, volatility, as well as news sentiment that we use for estimation. Notably, a drop in the S\&P 500 since mid-2015 is triggered by the Chinese stock market turbulence. One observes a simultaneous drop in the price and sentiment and surge in the intraday VIX. 
While news sentiment seems to be a realization of a typical mean-reverting Ornstein-Uhlenbeck process (with a very quick reverting speed) in which the overstated or understated news is promptly reverting to the level it should be. Our main goal is to find how it interacts with volatility in the model and impacts price indirectly via volatility process: 
\begin{align}
    \begin{split}
    dS_t &= \lambda_s (\mu_s - S_t) dt + \sigma_s dW_{s,t} \\
    dP_t &= (\mu_p - \exp(V_t)/2 ) dt +  \exp(V_t/2) dW_{p,t} \\
    dV_t &= (\mu_v + \beta_v (S_t - \mu_s)^2 - \gamma_v V_t) dt + \sigma_v dW_{v,t}
    \end{split}
\end{align}
where $corr(dW_{p,t},dW_{v,t})=\rho_{pv}$, $corr(dW_{s,t},dW_{v,t})=\rho_{sv}$. Note the process is stationary if $\mu_p = \exp(\mu_v/\gamma_v + \sigma_s^2/(2\lambda_s))/2$. We estimate parameters with $dt=1/26$ corresponding to 15-min interval of data. 
\begin{figure}[ht!]
\begin{center}
\includegraphics[width=\textwidth]{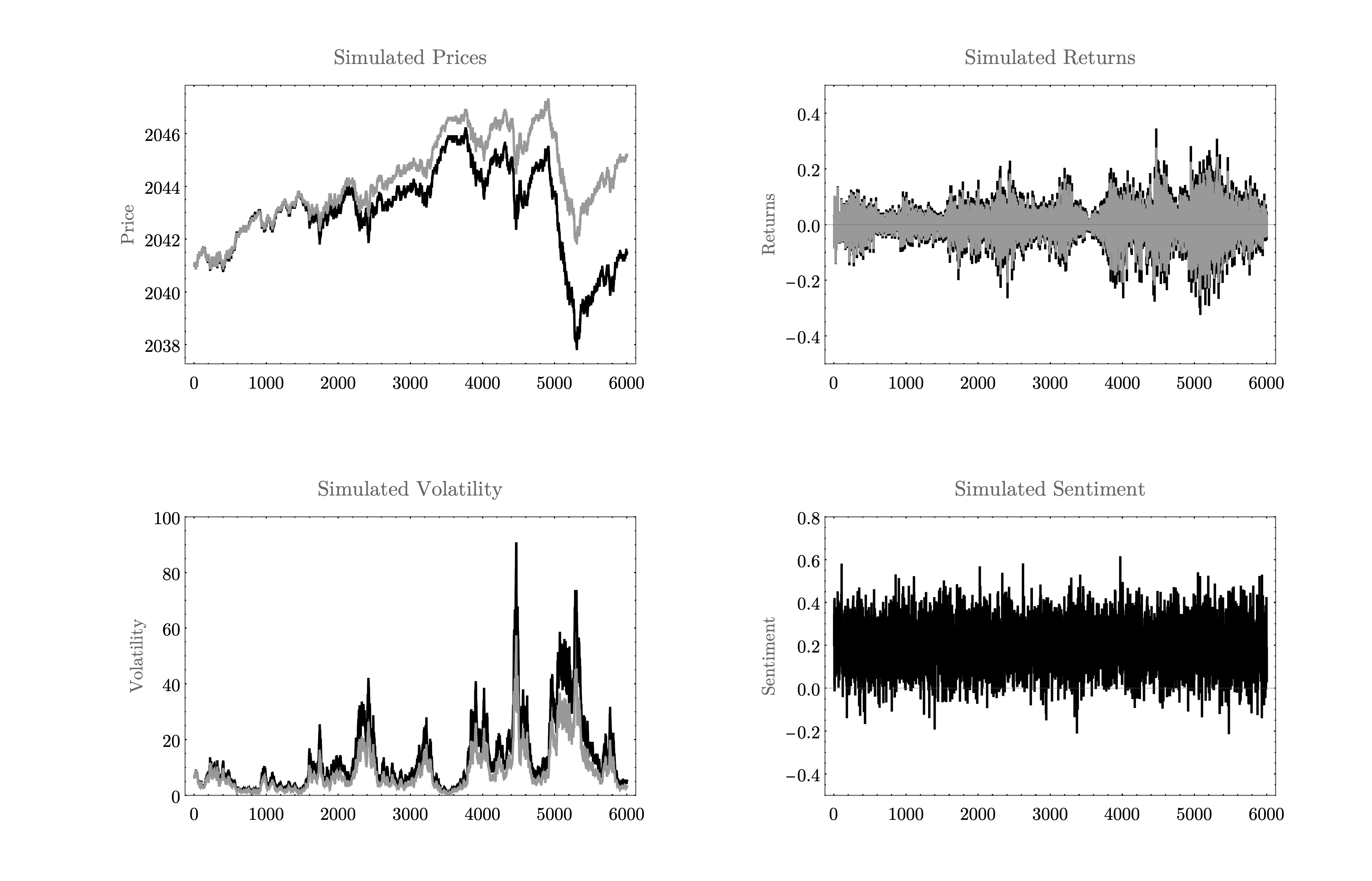}  \\
\caption{A realization of the sentiment-driven stochastic volatility using estimated parameters in black,  compared to the realization with $\beta_v = 0$ in gray.}
\label{fig:process}
\end{center}
\end{figure}
For the estimation, we implement NPSMLE using parallel \Cpp 11 code.\footnote{The code for estimation available on \url{https://github.com/barunik/npmsle}. We are grateful to Rastislav Kisel for research assistance and help with the implementation.} Number of simulations we use is 1000, and bandwidths in a multiplicative Gaussian kernel are chosen by the rule of thumb of \cite{scott2015multivariate}. Note we have also estimated parameters with variety of bandwidths without impact on results. These results are in line with findings of \cite{kristensen2012estimation} hence we do not report it here. 
\begin{table}
\label{tab:results}
\begin{center}
  \begin{tabular}{ l  rrrrrrrrrr }
    \toprule
    & $\lambda_s$ & $\mu_s$ &$\sigma_s$ & $\mu_p$& $\mu_v$& $\gamma_v$& $\beta_v$& $\sigma_v$& $\rho_{pv}$ & $\rho_{sv}$ \\ 
    \midrule
    true/simulated   &  37.76 & 0.203 & 0.916 & 0.0388 & -0.148 &  0.049  &  1.86  &  0.379 &  -0.89 & -0.025 \\ 
    \midrule
    mean   &  40.06 &  0.209 & 1.001        & 0.0534 & -0.154 & 0.061  & 2.075  & 0.384   & -0.915 & -0.03 \\
    median &  39.85 &  0.209 & 1.001        & 0.0421 & -0.164  & 0.062  & 2.234  & 0.384   & -0.915 & -0.03 \\
    std    &  1.278 &  0.001 & 0.0127        & 0.0806 & 0.0596  & 0.038  & 0.830  & 0.002  & 0.0025 & 0.045\\
    \bottomrule
  \end{tabular}
    \caption{Parameter estimates on the S\&P 500 data together with standard errors.}
\end{center}
\end{table}

Finally, we arrive with the set of parameter estimates in the frist row of Table \ref{tab:results}. $\lambda_s=37.76$ shows a very quick reversion, indicating that even there exists an exaggerated news tone or partially disclosed news, the following news will correct this deviation promptly. It seems that news has a very short journey, diffuses rapidly across the public, and will be absorbed by the market and reflected in the asset prices efficiently. In general, the news tones look optimistic with $\mu_s=0.203$. The higher volatility of turbulent markets can be sustained given the estimate of $\beta_v = 1.86$, as the level of reversion is elevated by a deviated news tone from its reasonable level. Apart from it, the value of $\E[V_t]$ is governed by $\beta_v$. $\rho_{pv}=-0.89$ supports the well-documented the instantaneous leverage effect in the literature.
$\rho_{vs}$, measuring the instantaneous correlation between two separate Brownian motions driving the sentiment and volatility process is -0.025. Although it's neither strong nor significant as $\rho_{pv}$, $\rho_{vs}$ may shed similar light as the leverage effect, implying a drop in sentiment an increase in the volatility. Note that $\rho_{vs}$ decides the quantity of $\E[S_t  V_t]$ in (11) and also the resulting correlation $\rho(S_t, V_t)$ in (19).

Figure \ref{fig:process} shows a realization of the simulated process from the estimated coefficients together with the restriction on $\beta_v = 0$, which is the process isolating an impact from news sentiment. One can observe that the model generates a very similar process as the real world data plotted by Figure \ref{fig:data}. Most importantly, the model incorporating the news effect generates higher volatility than the SV model. Through the simulated volatility, we document that an overstated or understated news tone sustains market volatility, prolonging the market turbulence.  

To be able to distinguish if the main parameter of interest, $\beta_v$ is statistically significant, we further estimate the bootstrapped standard errors from simulations. Table \ref{tab:results} show the mean, median and standard errors estimates from 1000 simulations with the same setting as in estimation. Here we generate the 1000 sample paths of the same length as data from the estimated parameters, and estimate the parameters on the generated data to obtain simulated inference. This small simulation study can be also used to see how well the estimation is able to determine the parameters. 

We can see that parameters have small bias, which is within the bias reported by \cite{kristensen2012estimation}. The high-frequency sentiment process is estimated very precisely, also $\sigma_v$, $\rho$, in volatility process are very precise. $\mu_p$ is not significant. According to stationarity condition, $\mu_p$ should be equal to 0.0325 which is true within standard error, so the estimation identifies parameters correctly. Parameters of volatility process, $\mu_v$ and $\gamma_v$ are estimated precisely as well, and most important, parameter $\beta_v$ connecting volatility to sentiment is significantly different from zero. 

\section{Conclusions}
We introduce textual analysis tools for quantifying high-frequency news sentiment to economists. News arriving randomly at markets and the resulting news sentiment behave like a stochastic process.  Due to exaggerated news tones or slow news diffusion, news may evolve like a mean-reverting Ornstein-Uhlenbeck process to converge to the level it should reach.  This OU-type news tone process may explain the well-documented investor's overreaction and underreaction in reality.

The sentiment effects on return and volatility are explored separately and haven't yet been formulated in a unified framework in the existing literature. To characterize the joint evolution of news sentiment, price, and volatility, we propose a continuous-time sentiment-driven stochastic volatility model. We study the theoretical properties of the proposed process and provide closed-form formulas for moments and co-moments of the sentiment and volatility. Such efforts enable us to measure how sentiment impacts volatility theoretically and analytically.
Employing the nonparametric simulated maximum likelihood estimation, we calibrate the parameters and use them for further simulation tasks. 

Empirically, we document interesting high-frequency behavior of news sentiment and volatility.  We find a very quick reversion in the news process, indicating that an exaggerated or understated news tone will be soon corrected by the following news. Besides, the news sentiment process is much less persistent than the volatility process, making a prediction challenge.  We may attribute long-lasting market turbulence to an elevated threshold of volatility reversions in the presence of news effects. We show how textual analytic tools introduce a brand new stochastic process into a continuous-time stochastic volatility framework. 


\section{Technical Appendix}
\label{sec:app}

Details of the Theorem \ref{th1}:

\begin{dmath}
a=\lambda_s^2 \left(\gamma_v^2+3 \gamma_v \lambda_s+2 \lambda_s^2\right) e^{4 \lambda_s t} \Bigl[2 \beta_v^2 \sigma_s^2 (2 \gamma_v-\lambda_s) \left(2 \gamma_v
   \mu_s^2-\sigma_s^2\right)+8 \beta_v \gamma_v \mu_s \rho_{sv} \sigma_s \sigma_v \left(\gamma_v^2-3 \gamma_v \lambda_s+2 \lambda_s^2\right)+\sigma_v^2 (\gamma_v-2 \lambda_s)^2 \left(2 \gamma_v^2-3 \gamma_v \lambda_s+\lambda_s^2\right)+4 \beta_v^2 \gamma_v S_0^2 \sigma_s^2 (2
   \gamma_v-\lambda_s)-8 \beta_v \gamma_v S_0 \sigma_s \left(\beta_v \mu_s \sigma_s (2 \gamma_v-\lambda_s)+\rho_{sv} \sigma_v \left(\gamma_v^2-3 \gamma_v \lambda_s+2 \lambda_s^2\right)\right)\Bigr]
\end{dmath}

\begin{dmath}
b= 2 \beta_v^2 \sigma_s^2 \left(2 \gamma_v^3+5 \gamma_v^2 \lambda_s+\gamma_v \lambda_s^2-2 \lambda_s^3\right) (\gamma_v-2 \lambda_s)^2 e^{2  (\gamma_v+\lambda_s)t} \left(2 \lambda_s \mu_s^2+2 \lambda_s S_0^2-4 \lambda_s \mu_s S_0-\sigma_s^2\right)
\end{dmath}

\begin{dmath}
c=\beta_v^2 \gamma_v \sigma_s^2 \left(2 \gamma_v^4+3 \gamma_v^3 \lambda_s-4 \gamma_v^2 \lambda_s^2-3 \gamma_v \lambda_s^3+2 \lambda_s^4\right)
   e^{2 \gamma_v t} \left(4 \lambda_s \mu_s^2+4 \lambda_s S_0^2-8 \lambda_s \mu_s S_0-\sigma_s^2\right)
\end{dmath}

\begin{dmath}
d=8 \beta_v \lambda_s^2 \sigma_s \left(2
   \gamma_v^2-3 \gamma_v \lambda_s+\lambda_s^2\right) e^{ (\gamma_v+2 \lambda_s)t} \Bigl[ 2 \beta_v \sigma_s \left(-\gamma_v^2 \mu_s^2+\gamma_v
   \left(\sigma_s^2-3 \lambda_s \mu_s^2\right)+\lambda_s \left(\sigma_s^2-2 \lambda_s \mu_s^2\right)\right)-\gamma_v \mu_s \rho_{sv} \sigma_v
   \left(\gamma_v^2-4 \lambda_s^2\right)-2 \beta_v S_0^2 \sigma_s \left(\gamma_v^2+3 \gamma_v \lambda_s+2 \lambda_s^2\right)+S_0 (\gamma_v+2 \lambda_s) (4
   \beta_v \mu_s \sigma_s (\gamma_v+\lambda_s)+\gamma_v \rho_{sv} \sigma_v (\gamma_v-2 \lambda_s))\Bigr] 
\end{dmath}

\begin{dmath}
e=8 \beta_v \gamma_v \lambda_s^2
   \rho_{sv} \sigma_s \sigma_v \left(\gamma_v^2+\gamma_v \lambda_s-2 \lambda_s^2\right) (\gamma_v-2 \lambda_s)^2 (S_0-\mu_s) e^{(2 \gamma_v +3
   \lambda_s) t}
\end{dmath}

\begin{dmath}
f=\left(2 \gamma_v^3-\gamma_v^2 \lambda_s-2 \gamma_v \lambda_s^2+\lambda_s^3\right) (\gamma_v-2 \lambda_s)^2 e^{2  (\gamma_v+2 \lambda_s)t}
   \left(\beta_v^2 \sigma_s^4+\lambda_s^2 \sigma_v^2 (\gamma_v+2 \lambda_s)\right)
\end{dmath}

Proof of the Proposition \ref{S}:

\begin{proof}
This follows from the fact that the process $S_t$ is an Ornstein-Uhlenbeck process. It has normal distribution and
admits a solution
\begin{equation}
S_t = e^{-\lambda_s t} S_0 + \mu_s[1 - e^{-\lambda_s t}] + \int_0^t \sigma_s   e^{\lambda_s (u-t)} dW_u^s.
\end{equation}
\end{proof}

Proof of the Proposition \ref{V}:

\begin{proof}
Multiplication of $V_t$ by an integrating factor $e^{\gamma_v t}$ removes the $V_t$ term in the right side of  
equation (\ref{V}):
\begin{equation}
d( e^{\gamma_v t} V_t) = e^{\gamma_v t} (\mu_v + \beta_v(S_t - \mu_s)^2) dt  + e^{\gamma_v t}  \sigma_v dW_t^v.
\end{equation}
The result of the proposition follows from integration and taking the expectation.
\end{proof}

Proof of the Proposition \ref{SV}:

\begin{proof}
Ito's formula for the product gives
\begin{eqnarray}
d(S_t V_t) &=& S_t dV_t + V_t dS_t + dS_t dV_t \\
&=& S_t (\mu_v + \beta_v(S_t - \mu_s)^2 - \gamma_v V_t) dt +  \sigma_v S_t dW_t^v \notag \\ 
&& + \lambda_s (\mu_s - S_t) V_t dt + \sigma_s V_t dW_t^s + \sigma_s \sigma_v \rho_{sv} dt. \notag
\end{eqnarray}
After integration, taking the expectation and use of $e^{(\gamma_v + \lambda_s)t}$ as an integrating factor, 
we get the result listed in the proposition.
\end{proof}

Proof of the Proposition \ref{EV2}:

\begin{proof}
From Ito's formula, we get
\begin{eqnarray*}
 d (e^{2\gamma_v t}V_t^2) &=& 2 \gamma_v e^{2\gamma_v t}V_t^2 dt + e^{2\gamma_v t} dV^2_t \\
 &=&
 2 \gamma_v e^{2\gamma_v t}V_t^2 dt+ 2e^{2\gamma_v t} V_t dV_t + e^{2\gamma_v t} (dV_t)^2 \\
 &=& 2V_t e^{2\gamma_v t}(\mu_v + \beta_v(S_t - \mu_s)^2) dt + \sigma_v e^{2\gamma_v t} V_t  dW_t^v + 
 e^{2\gamma_v t}\sigma_v^2 dt.
\end{eqnarray*}
The statement follows from integration and taking the expectation.
 \end{proof}

Proof of the Proposition \ref{ES2V}:

\begin{proof}
From Ito's formula, we get
\begin{eqnarray*}
d(S_t^2 V_t) &=& S_t^2 dV_t + V_t dS_t^2 + dV_t dS_t^2 \\
&=& S_t^2 (\mu_v + \beta_v(S_t - \mu_s)^2 - \gamma_v V_t) dt + \sigma_v S_t^2 dW_t^v \\
&& + 2 S_t V_t (\lambda_s (\mu_s - S_t) dt + \sigma_s dW_t^s) + \sigma_s^2 V_t dt + 2 S_t \sigma_v \sigma_s \rho_{sv}dt
\end{eqnarray*}
The statement follows from integration and taking the expectation.
\end{proof}


{\footnotesize
\bibliographystyle{dcu}
\bibliography{bibJabRef}
}



\end{document}